\documentclass[]{spie}  

\usepackage{amsmath,amsfonts,amssymb}
\usepackage{graphicx}
\usepackage{booktabs}
\usepackage[colorlinks=true, allcolors=blue]{hyperref}
\usepackage{tabularx}
\newcolumntype{C}[1]{>{\centering\arraybackslash}p{#1}}

\usepackage[final]{microtype}

\usepackage[dvipsnames]{xcolor}
\usepackage{fontawesome}
\usepackage[outline, auto]{contour}
\contourlength{0.50pt}
\newcommand{\tick}{\contour{Gray}{\textcolor{Green}{\faCheck}}}
\newcommand{\xmark}{\contour{Gray}{\textcolor{Maroon}{\faTimes}}}

\title{A latent space for unsupervised MR image\\ quality control via artifact assessment}

\author[a,b]{Lianrui~Zuo}
\author[a]{Yuan~Xue}
\author[c]{Blake~E.~Dewey}
\author[a]{Yihao Liu}
\author[a]{\\Jerry~L.~Prince}
\author[a]{Aaron~Carass}
\affil[a]{Department of Electrical and Computer Engineering, Johns~Hopkins~University,~Baltimore,~MD~21218,~USA}
\affil[ ]{}
\affil[b]{Laboratory of Behavioral Neuroscience, National Institute on Aging, National~Institutes~of~Health,~Baltimore,~MD~20892,~USA}
\affil[ ]{}
\affil[c]{Department of Neurology, Johns~Hopkins~School~of~Medicine, Baltimore,~MD~21287,~USA}

\authorinfo{Corresponding author: Lianrui Zuo. \\
Email: \texttt{lr\_zuo@jhu.edu}}

\begin{document} 
\maketitle

\begin{abstract}
Image quality control~(IQC) can be used in automated magnetic resonance~(MR) image analysis to exclude erroneous results caused by poorly acquired or artifact-laden images.
Existing IQC methods for MR imaging generally require human effort to craft meaningful features or label large datasets for supervised training.
The involvement of human labor can be burdensome and biased, as labeling MR images based on their quality is a subjective task.
In this paper, we propose an automatic IQC method that evaluates the extent of artifacts in MR images without supervision. 
In particular, we design an artifact encoding network that learns representations of artifacts based on contrastive learning.
We then use a normalizing flow to estimate the density of learned representations for unsupervised classification. 
Our experiments on large-scale multi-cohort MR datasets show that the proposed method accurately detects images with high levels of artifacts, which can inform downstream analysis tasks about potentially flawed data.

\end{abstract}


\section{INTRODUCTION} 
\label{sec:intro}
The recent development of deep learning~(DL) has benefited various magnetic resonance~(MR) image analyses, such as image synthesis~\cite{zuo2020synthesizing, zuo2022disentangling}, segmentation~\cite{huo20193d, liu2022disentangled}, registration~\cite{liu2022coordinate}, and volumetric analysis~\cite{duan2022cranial}, where a large amount of images are processed without human intervention. 
Yet, these DL based algorithms are known to be sensitive to the quality of input images~\cite{kugler2018dlf}; when an image is poorly acquired or contaminated by artifacts, the DL algorithms are likely to produce erroneous or biased results. 
Manually inspecting DL results in large datasets is prone to errors as it is tedious and subjective. 
Therefore, there is demand for an automatic image quality control~(IQC) method to identify potential failures cases caused by either \underline{poor quality} or \underline{inappropriate data}.

Various IQC methods have been developed in recent years~\cite{esteban2017mriqc,kang2014convolutional}.
The goal of an IQC method is to provide an assessment of image quality~$\hat{y}$ based on the input image~$x$.
In general, an IQC algorithm has two parts: feature extraction and classification. 
Features $m$ that capture image quality information can either be handcrafted with expert knowledge~\cite{esteban2017mriqc,zuo2018automatic} or learned from data~\cite{kang2014convolutional}.
Classification is conducted based on the features $m$, which usually requires expert labels on a sample dataset---e.g., with $y \in \{0,1\}$ indicating whether image $x$ passes or fails quality inspection---from which a supervised classifier is trained.
For example, MRIQC~\cite{esteban2017mriqc} learned a binary classifier based on handcrafted features and labels generated by human experts. 
However, the current IQC methods face two major limitations.
First, labeling datasets by experts requires domain specific knowledge, which can be subjective and time consuming.
Second, because the labels $y$ are limited in number and dataset specific, current IQC methods usually have limited generalizability. 
The feature extractor and classifier are usually generalizable to datasets similar to what they have been trained on; however, new datasets will generally require a re-training or fine tuning, critically this necessitates new labels. 

To overcome the limitations of current IQC methods, we developed an unsupervised IQC method to directly assess artifact levels from MR images. 
Our method has two advantages. 
First, we propose an artifact encoder network that learns latent artifact representations in a data-driven way.
Second, we use a normalizing flow~\cite{dinh2016density} to map the learned representations $m$ to a normal distribution, which allows us to conduct unsupervised classification without expert labels $y$. 
It is worth noting that the artifact encoder is also unsupervised, meaning that no labels are needed in our method.
This means our approach is completely unsupervised, making our framework applicable to more datasets. 

\section{METHODS}
\label{sec:methods}
\begin{figure}[!tb]
    \centering
    \includegraphics[width=0.75\textwidth]{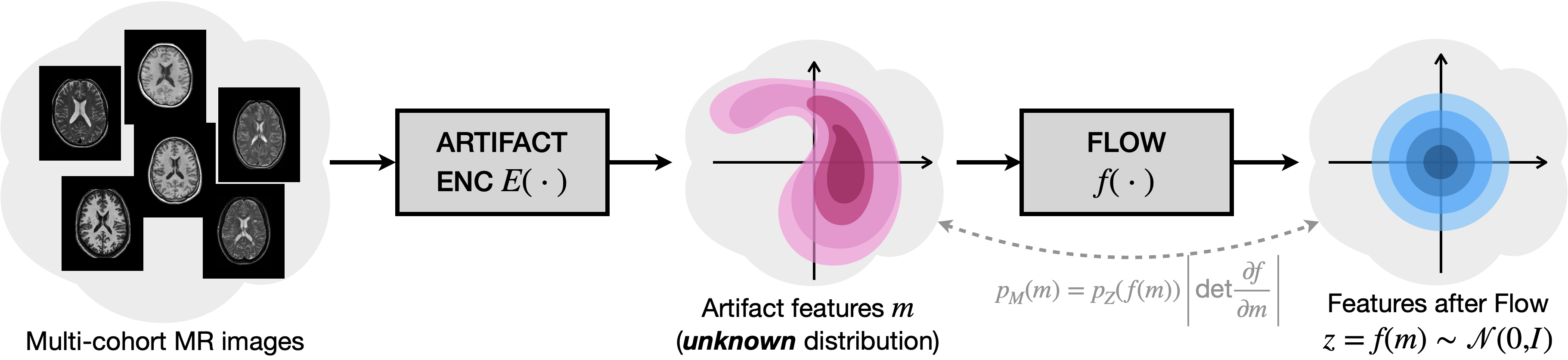}
    \caption{Schematic framework of the proposed IQC method. Artifact encoder $E(\cdot)$ extracts artifact features $m \in \mathbb{R}^2$ from multi-cohort MR images. The learned features $m$ follow an unknown distribution $p_M(m)$. A normalizing flow $f(\cdot)$ is then applied to transform $m$ to $z \in \mathbb{R}^2$ following a standard Gaussian distribution $\mathcal{N}(0,I)$. Due to the special property of $f(\cdot)$, the likelihood $p_M(m)$ can be evaluated using Eq.~\ref{eq:density}.}
    \label{fig:framework}
\end{figure}

Figure~\ref{fig:framework} shows the framework of the proposed method. MR images from multiple cohorts are first encoded into a two-dimensional latent space of artifact features $m$. In general, $m$ follows an unknown distribution $p_M(m)$. We then apply a normalizing flow~\cite{dinh2016density} $f(\cdot)$ to transform $m$ to $z\in \mathbb{R}^2$, which follows a standard Gaussian distribution. $f(\cdot)$ also enables density estimation of $p_M(m)$ for unsupervised IQC.

\subsection{Artifact encoder based on contrastive learning}\label{sec:encoder}
\begin{figure}[!tb]
    \centering
    \begin{tabular}{c c}
        \includegraphics[width=0.61\textwidth]{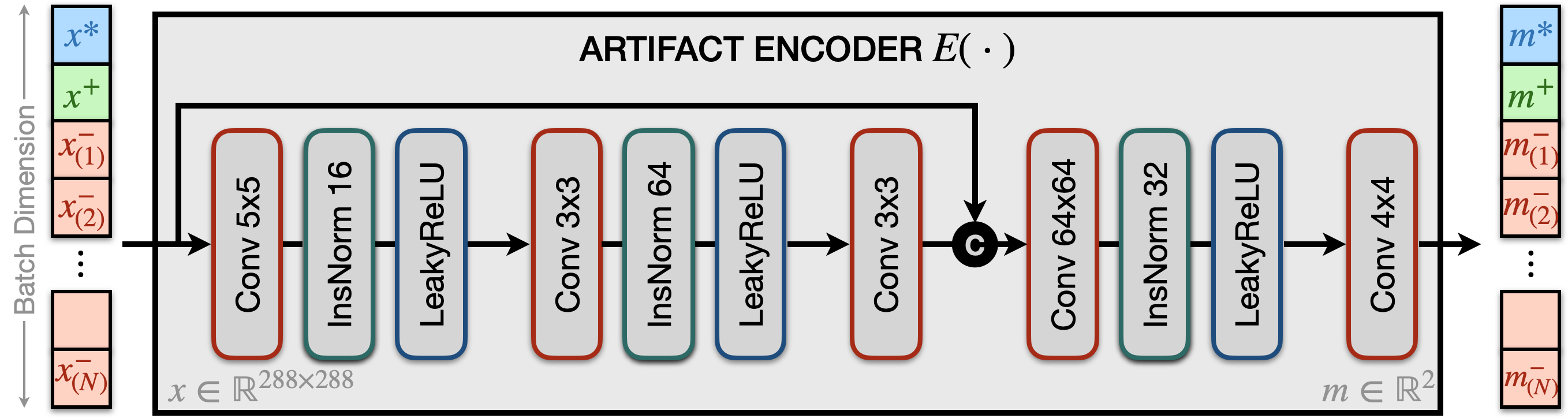}  &  
        \includegraphics[width=0.33\textwidth]{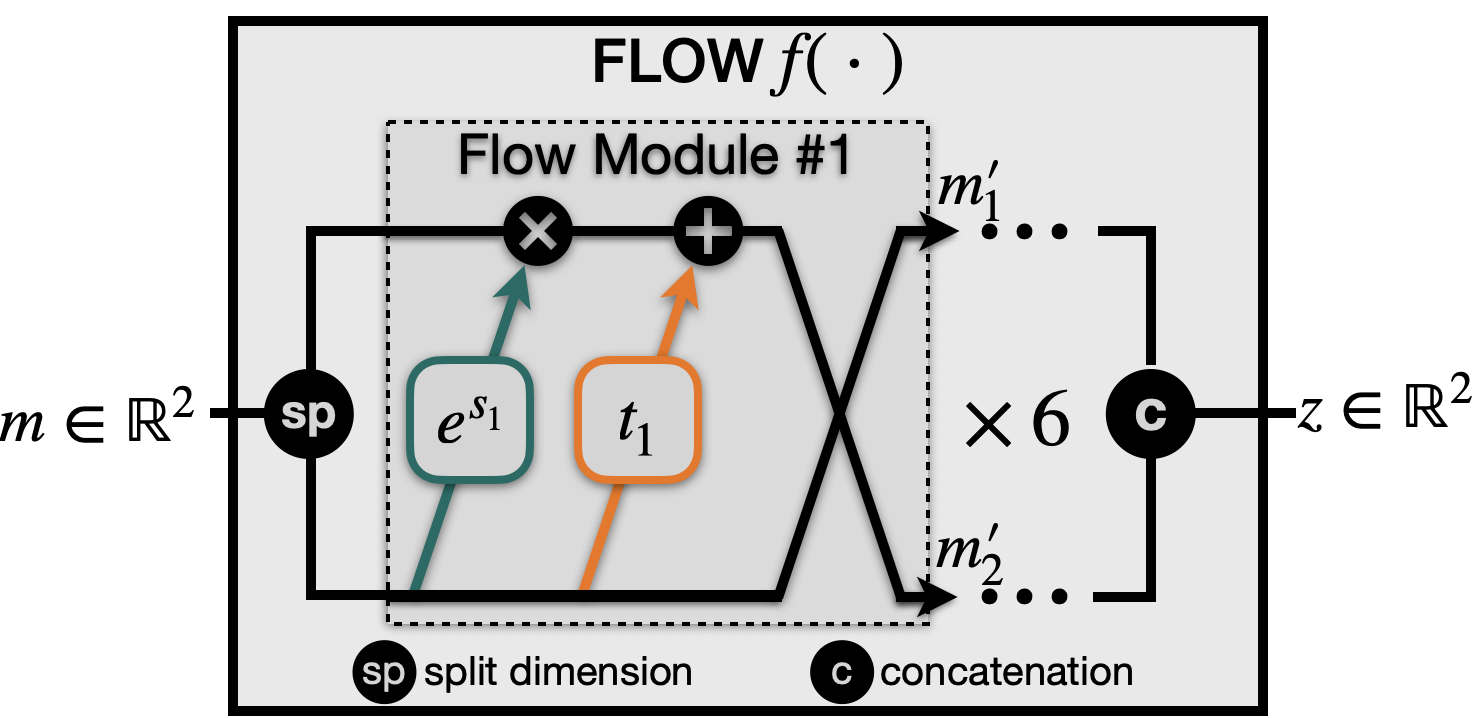}  \\
        \textbf{(a)} & \textbf{(b)}
    \end{tabular}
    \caption{\textbf{(a)} The artifact encoder has a Dense-Net architecture. Artifact representations $m$ are learned based on contrastive learning with both positive images $x^+$ and negative images $x^-_{(i)}$. $x^*$ shares the same and different artifact levels with $x^+$ and $x^-_{(i)}$, respectively. \textbf{(b)} The flow network is based on RealNVP~\cite{dinh2016density} with six flow modules (affine coupling layers). $\{s_i, t_i \}_{i=1}^6$ are trainable neural networks. }
    \label{fig:networks}
\end{figure}

Our artifact encoder $E(\cdot)$ extracts artifact representations based on contrastive learning~\cite{park2020contrastive}. 
The key concept of contrastive learning is to learn discriminative features from query, positive, and negative examples. 
Figure~\ref{fig:networks}(a) shows the architecture of $E(\cdot)$. 
For each MR image $x^*\in \mathbb{R}^{288\times288}$, we assume the positive example $x^+$ has the same artifact level as the query example $x^*$.
We achieve this by selecting image slices $x^*$ and $x^+$ from different orientations of the same 3D volume~(e.g., axial and coronal slices). 
Our negative examples $\{x^-_{(i)}\}_{i=1}^N$ are chosen to have different artifact levels than $x^*$. 
We prepare our negative examples by either selecting slices from a volume different from the source of $x^*$ or augmenting $x^*$ with simulated artifacts including noise and motion. 
The simulated images are used to prevent $E(\cdot)$ from learning irrelevant information such as contrast and anatomy, since slices from different volumes may differ both in their level of artifacts and in their contrasts and anatomies.
Because we also introduce real MR images as negative examples, $E(\cdot)$ after training can capture different kinds of artifacts---beyond just noise and motion---which we show in Sec.~\ref{sec:experiments}.

With $x^*$, $x^+$, and $x^-_{(i)}$'s composing our input mini-batch, we expect the learned feature $m^*$ to be similar (if not identical) to $m^+$ and sufficiently distinct from the $m^-_{(i)}$'s. We encourage this relationship using 
\begin{equation}
    \mathcal{L}(m^*, m^+, \{ m^-_{(i)} \}_{i = 1}^N) = -\log \left[ \frac{\exp(m^* \cdot m^+)}{\exp(m^* \cdot m^+) + \frac{1}{N}\sum_{i = 1}^N \exp(m^* \cdot m^-_{(i)})}  \right]
    \label{eq:contrastive_learning}
\end{equation}
as our loss function for $E(\cdot)$. 
Since we prepare our $x^+$ and $x^-_{(i)}$ based on their relative extent of artifact with $x^*$ and encourage $m$ to preserve this relationship, we would expect $m$ to capture the artifact information of the input image. 
Note that $m$ is learned based on the \textit{relative} extent of artifact between $x^*$, $x^+$, and $x^-_{(i)}$, there is no assumption made about the \textit{absolute} extent of artifact of $x^*$ (i.e. $x^*$ is not assumed to be free from artifacts).

\subsection{Density estimation with normalizing flows}\label{sec:flow}
With sufficiently large datasets, one can assume that most acquired MR images have acceptable image quality with a relatively small sample of images being poorly acquired or contaminated by artifacts. 
This ratio is reflected by the likelihood $p_M(m)$; when an image has uncommonly high artifact level $\tilde{m}$, we would expect $p_M(\tilde{m})$ to be small. 
Unsupervised IQC can then be achieved by finding $\{x_i\}$'s with $p_M(m_i)$'s below a percentile. Evaluating $p_M(m)$ is a nontrivial task, but it can be approximated using a normalizing flow network~\cite{dinh2016density} $f(\cdot)$. 
As shown in Fig.~\ref{fig:networks}(b), $f(\cdot)$ is composed of six flow modules with each module affinely processing a proportion of the input variable, e.g., $m_1' = m_2$ and $m_2' = m_1 \cdot e^{s_1(m_2)} + t_1(m_2)$, where $m=[m_1,m_2]\in \mathbb{R}^2$ and $\{s_i, t_i \}_{i=1}^6$ are neural networks.
The output variable $z=f(m)$ follows a standard Gaussian distribution $\mathcal{N}(0,I)$. 
It is easy to show that the Jacobian matrix of $f(m)$ is a triangular matrix with positive determinant and the density $p_M(m)$ can be calculated by
\begin{equation}
    p_M(m) = p_Z\left( f\left( m \right) \right) \left| \text{det} \frac{\partial f(m)}{\partial m} \right|.
    \label{eq:density}
\end{equation}
During training, we use $-\log p_M(m)$ calculated with Eq.~\ref{eq:density} as our loss function for $f(\cdot)$, where the trainable modules are $\{s_i, t_i \}_{i=1}^6$. 

\section{EXPERIMENTS AND RESULTS}
\label{sec:experiments}
\subsection{Datasets and preprocessing}
The training data for $E(\cdot)$ include 200 T$_1$-weighted~(T$_1$-w) MR volumes acquired from $16$ different cohorts. 
Detailed information about image acquisition is provided in Table~\ref{tab:data}.
Our preprocessing includes inhomogeneity correction~\cite{tustison2010n4itk} and registration to a $0.8$mm$^3$ isotropic template.
For each 3D volume, we extracted and padded axial, coronal, and sagittal slices to dimension $288 \times 288$.
$E(\cdot)$ was trained on 2D slices following Sec.~\ref{sec:encoder}.
Our evaluation dataset for $E(\cdot)$ has 1,400 3D images acquired from the $16$ cohorts. 
For each volume, we calculated the average $m$ values of its 20 center axial slices as the artifact representation.
$f(\cdot)$ was then trained and applied following Sec.~\ref{sec:flow} to estimate the density $p_M(m)$ based on all the 1,400 volumes.

\begin{table}[!tb]
    \centering
    \caption{Key information about image acquisition of each imaging cohort. Unavailable information is marked as ``--''. Data source: $C_1$ and $C_2$~(IXI-Brain)~\cite{ixi}; $C_3$ thru $C_6$~(OASIS3)~\cite{LaMontagne}; $C_7$ thru $C_{10}$~(BLSA)~\cite{resnick2000one}; $C_{11}$ thru $C_{16}$~(Private).}
    \resizebox{0.97\columnwidth}{!}{
    \begin{tabular}{C{0.15\columnwidth} 
                    C{0.12\columnwidth} C{0.12\columnwidth} C{0.12\columnwidth} C{0.12\columnwidth}
                    C{0.12\columnwidth} C{0.12\columnwidth} C{0.12\columnwidth} C{0.12\columnwidth}
                    }    
    \\[-0.6em]
    \toprule
    {\bf Cohort} & {$C_1$} & {$C_2$} & {$C_3$} & {$C_4$} & {$C_5$} & {$C_6$} & {$C_7$} & {$C_8$} \\
    \cmidrule(lr){2-9}
    {\bf Open data} & \tick & \tick & \tick & \tick & \tick & \tick & \tick & \tick \\
    \cmidrule(lr){2-9} 
    {\bf Manufacturer} & Philips & Philips & Siemens & Siemens & Siemens & Siemens & Philips & Philips \\
    \cmidrule(lr){2-9}
    {\bf Field~(T)} & $1.5$ & $3.0$ & $3.0$ & $3.0$ & $3.0$ & $1.5$ & $1.5$ & $3.0$ \\
    \cmidrule(lr){2-9}
    {\bf Resolution~(mm)} & $1.2\times0.9\times0.9$ & $1.2\times0.9\times0.9$ 
                        & $1.0\times1.0\times1.0$ & $1.0\times1.0\times1.0$ 
                        & $1.0\times1.0\times1.0$ & $1.1\times1.1\times1.2$ 
                        & $0.9\times0.9\times1.5$ & $1.0\times1.0\times1.2$ \\
    \cmidrule(lr){2-9} 
    {\bf TE/TR/TI~(ms)} & $4.6/-/-$ & $4.6/-/-$ & $3.9/1900/1100$ 
                        & $3.2/2400/1000$ & $3.2/2400/1000$ & $2.9/2300/900$ & $3.3/3000/-$ 
                        & $3.1/3000/800$\\
    \midrule
    {\bf Cohort} & {$C_9$} & {$C_{10}$} & {$C_{11}$} & {$C_{12}$} & {$C_{13}$} & {$C_{14}$} & {$C_{15}$} & {$C_{16}$} \\
    \cmidrule(lr){2-9}
    {\bf Open data} & \tick & \tick & \xmark & \xmark & \xmark & \xmark & \xmark & \xmark \\
    \cmidrule(lr){2-9} 
    {\bf Manufacturer} & Philips & Philips & Siemens & GE & Siemens & GE & Siemens & Siemens \\
    \cmidrule(lr){2-9}
    {\bf Field~(T)} & $1.5$ & $3.0$ & $3.0$ & $3.0$ & $3.0$ & $1.5$ & $1.5$ & $3.0$ \\
    \cmidrule(lr){2-9} 
    {\bf Resolution~(mm)} & $1.2\times0.9\times0.9$ & $1.2\times0.9\times0.9$ 
                        & $1.0\times1.0\times1.0$ & $1.0\times1.0\times1.0$ 
                        & $1.0\times1.0\times1.0$ & $1.1\times1.1\times1.2$ 
                        & $0.9\times0.9\times1.5$ & $1.0\times1.0\times1.2$ \\
    \cmidrule(lr){2-9}
    {\bf TE/TR/TI~(ms)} & $3.1/3000/800$ & $3.1/3000/800$ & $3.0/2300/900$ 
                        & $3.1/-/-$ & $3.6/2500/-$ & $2.6/-/-$ 
                        & $3.0/2300/900$ & $3.4/2300/900$\\
    \bottomrule
    \\[-0.5em]
    \end{tabular}}
    \label{tab:data}
\end{table}

\subsection{Unsupervised IQC on simulated data}
\label{sec:simulated_data}
\begin{figure}
    \centering
        \includegraphics[width=0.99\textwidth]{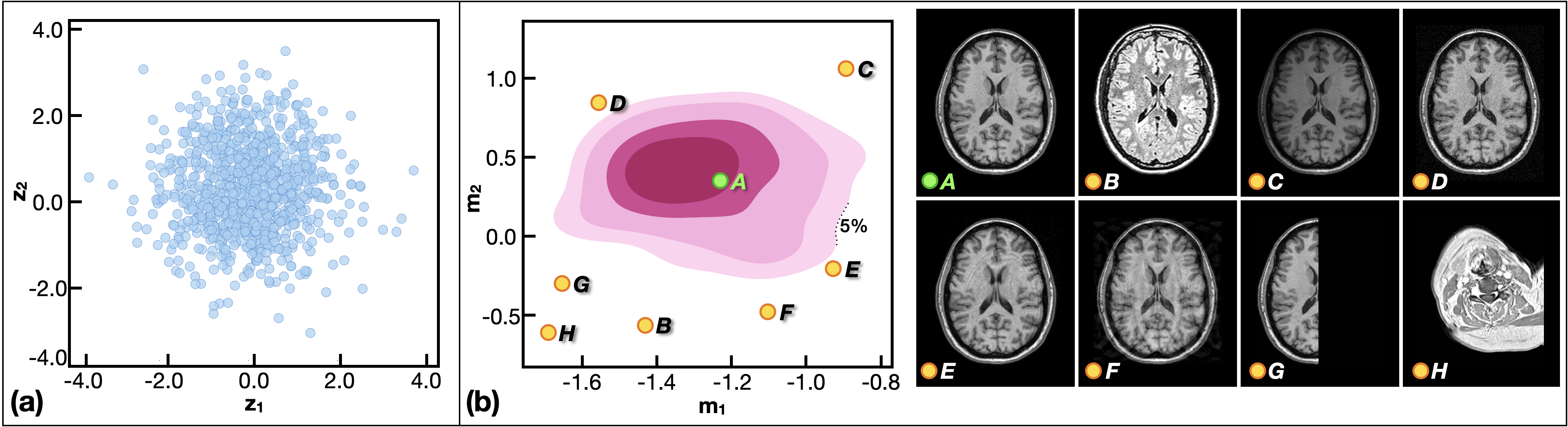}
    \caption{\textbf{(a)} scatter plot of $z$ from 1,400 T$_1$-w MR volumes after a normalizing flow $z = f(m)$. \textbf{(b)} $m$ values of held-out MR images are shown on top of the density contours fitted on $m$. Eight example MR images are shown on the right.
    Indexes $A$ to $H$ represent an MR image $A$)~a T$_1$-w image that passed our manual inspection, $B$)~a non-T$_1$-w image, $C$)~a T$_1$-w image with a bias field, $D$)~a T$_1$-w image with high noise, $E$)~a T$_1$-w image with motion artifacts, $F$)~a T$_1$-w image with wrap-around artifacts, $G$)~a T$_1$-w image with one side of the head removed, and $H$)~a T$_1$-w image with registration errors.}
    \label{fig:results}
\end{figure}

After training, $f(\cdot)$ transforms $m$ to $z$, which follows a standard Gaussian distribution.
Density $p_M(m)$ can then be evaluated using Eq.~\ref{eq:density}.
Figure~\ref{fig:results}(a) shows $z$ values of the 1,400 volumes after applying the normalizing flow $f(\cdot)$. 
We then applied the proposed method to a held-out simulated dataset with various kinds of artifacts that could potentially fail downstream analyses.
In Fig.~\ref{fig:results}(b), $m$'s of eight representative images are shown on top of the density contours of the 1,400 volumes.
The eight images are $A$)~a T$_1$-w image that passed our manual inspection, $B$)~a non-T$_1$-w image, $C$)~a T$_1$-w image with a bias field, $D$)~a T$_1$-w image with high noise, $E$)~a T$_1$-w image with motion artifacts, $F$)~a T$_1$-w image with wrap-around artifacts, $G$)~a T$_1$-w image with one side of the head removed, and $H$)~a T$_1$-w image with registration errors.
We assume the original 1,400 volumes are fairly diverse samples of T$_1$-w MR images and that most of them have acceptable artifact levels with a small proportion being poorly acquired. 
Unsupervised IQC is achieved by thresholding $p_M(m)$ with a predefined threshold $\tau$.
We found $\tau = 5\%$ achieved satisfactory results on our simulated dataset.
As shown in Fig.~\ref{fig:results}(b), the image that passed our manual inspection has an $m$ located in  the high density region, while the remaining seven images have $p_M(m)$ below $\tau$.
Our unsupervised IQC method has two advantages over existing works.
First, we do not require knowledge of the absolute artifact levels of training images, so that our method can be trained on very large datasets.
In fact, we only assume that most our training data have acceptable image quality; this is likely true in many application scenarios. 
Based on contrastive learning, our artifact encoder $E(\cdot)$ during training only needs to know if a sample has the same (for positive examples) or a different (for negative examples) artifact level as the query image $x^*$. 
Second, since we construct our negative examples with both real data and simulated artifacts~(i.e., motion and noise), $E(\cdot)$ after training can capture various kinds of artifacts, many of which have not been simulated in training.
This makes our model more generalizable. 

\subsection{Quantitative evaluation on real data}
\label{sec:real_data}
\begin{figure}[!tb]
    \centering
    \includegraphics[width=0.99\textwidth]{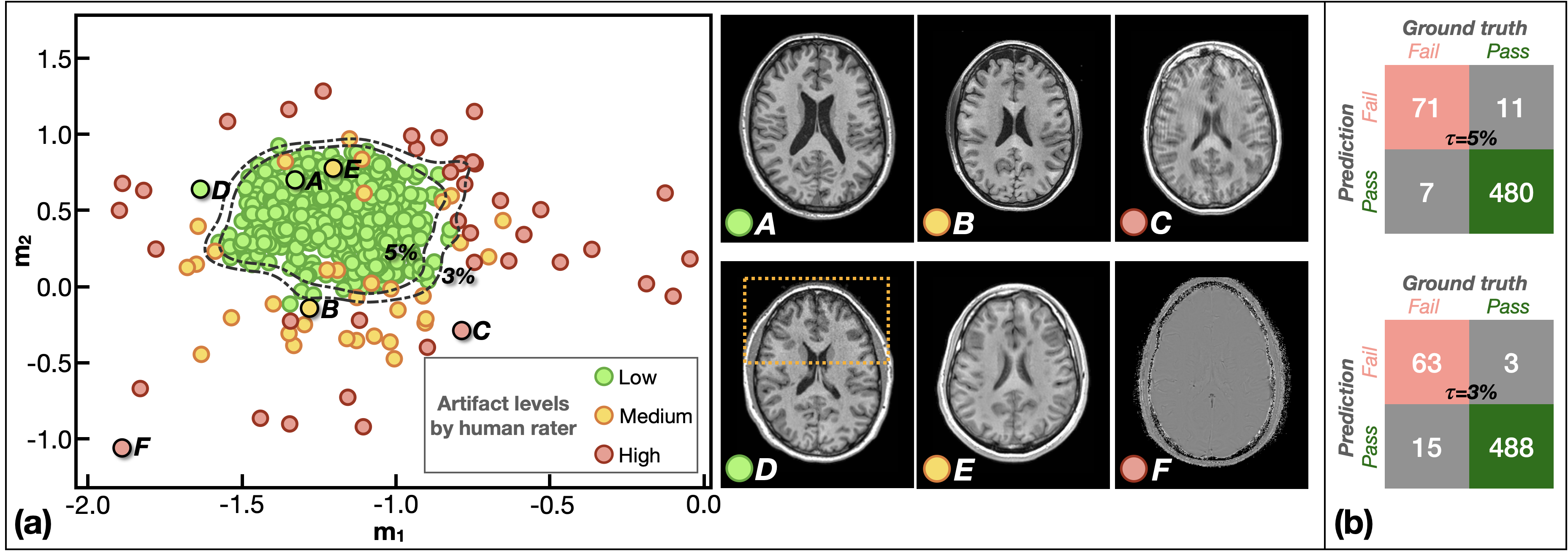}
    \caption{\textbf{(a)} Scatter plot of $m$ values on a dataset with manual ratings. Example images are shown on the right. Dashed lines show $5\%$ and $3\%$ likelihood contours of $p_M(m)$. \textbf{(b)} Contingency tables of the proposed method based on $\tau = 5\%$ and $\tau = 3\%$. Images with either medium or high artifacts levels are categorized as failed cases. }
    \label{fig:scatter_plot_real}
\end{figure}

To quantitatively evaluate the proposed method on real MR datasets, we manually inspected and rated $569$ T$_1$-w MR images acquired from cohorts $C_{11}$ to $C_{16}$~(see Table~\ref{tab:data} for more details). 
After manual inspection, each image was assigned a label from one of the three labels low, medium, or high based on the level of artifacts present in the volume.
We assume images with low levels of artifacts passed our manual quality check, and assume images with either medium or high artifact levels as failed cases.
Figure~\ref{fig:scatter_plot_real}(a) shows the learned $m$ values of the images with manual ratings.
Green, orange, and red represent low, medium, and high levels of artifacts, respectively.
Two density contours ($5\%$ and $3\%$) of $p_M(m)$ are also shown in Fig.~\ref{fig:scatter_plot_real}(a).
It is encouraging to see that most images that passed our manual inspection~(green) have $m$ values with $p_M(m) > 5\%$, while most images that failed our manual inspection (with medium and high artifact levels) have $p_M(m) < 5\%$.
Furthermore, images with high levels of artifacts~(red) usually have even lower $p_M(m)$ than images with medium levels of artifacts~(orange).
Figure~\ref{fig:scatter_plot_real}(a) also shows six example images with different levels of artifacts, where $A$)~has passed our manual quality check and it has $p_M(m) > 5\%$. 
$B$)~has a medium level of artifacts due to the intensity inhomogeneity, and 
$C$)~has strong motion artifacts. 
Interestingly, image $D$)~has passed our manual quality check, but the proposed method identified it as a low density example~($p_M(m) < 5\%$).
We hypothesize the reason for this is because the uncommon noise pattern of the image; the noise level is only high inside the orange box.
$E$)~shows an example with medium artifact level according to our manual inspections, but our method failed to identify it as a poor quality image. 
$F$)~is an extreme case where a non-T$_1$-w image was processed and identified by our algorithm as potential bad data.

In Fig.~\ref{fig:scatter_plot_real}(b), we show the contingency tables of the proposed method based on two thresholds: $\tau=5\%$ and $\tau=3\%$. 
Here, we assume any images with $p_M(m) < \tau$ at test time should be highlighted as potential artifact-laden images~(potential failed cases). 
$\tau=5\%$, which we used on simulated data in Sec.~\ref{sec:simulated_data}, achieves a sensitivity of $91.0\%$ and a specificity of $97.8\%$. 
$\tau=3\%$ achieves a sensitivity of $80.0\%$ and a specificity of $99.4\%$.

\section{Discussion and Conclusion}
\label{sec:conclution}
In this paper, we present a novel unsupervised IQC approach by assessing the levels of artifacts from MR images.
Our approach learns representations of image artifacts without domain knowledge.
This unsupervised nature enables our approach to be trained on a large variety of datasets with improved applicability over existing IQC methods. 
We showcase using normalizing flow that after artifact representations are learned, classification can be achieved with a simple thresholding on feature densities. 
The fact that the threshold $\tau$ needs to be determined at test time is a limitation of our work, as it may vary from dataset to dataset.
We regard this as a direction for future improvements. 
We believe introducing a very small amount of labels during training (for semi-supervised training) would benefit the proposed method to learn more robust feature extractors and classifiers. 

Experiments on both simulated and real MR datasets show that the proposed method achieves both high sensitivity and specificity. 
Our approach can be used to inform downstream analyses about potential bad quality data by accurately highlighting different kinds of artifact cases as low likelihood examples. 
 
\acknowledgments
This work was supported in part by the Intramural Research Program of the NIH, National Institute on Aging and in part by the TREAT-MS study funded by the Patient-Centered Outcomes Research Institute~(PCORI/MS-1610-37115).

\bibliography{report} 
\bibliographystyle{spiebib} 

\end{document}